\documentclass[twocolumn,showpacs,preprintnumbers,amsmath,amssymb]{revtex4}

\usepackage{graphicx}
\usepackage{dcolumn}
\usepackage{bm}

\usepackage{amssymb}
\usepackage{amsmath}

\begin{document}


\title{Horizon thermodynamics and cosmological equations: A holographic-like connection between thermostatistical quantities on a cosmological horizon and in the bulk}

\author{Nobuyoshi {\sc Komatsu}}  \altaffiliation{E-mail: komatsu@se.kanazawa-u.ac.jp} 
\affiliation{Department of Mechanical Systems Engineering, Kanazawa University, Kakuma-machi, Kanazawa, Ishikawa 920-1192, Japan}


\begin{abstract}

Horizon thermodynamics is expected to be related to the effective energy based on the energy density calculated from the Friedmann equation for a Friedmann--Robertson--Walker (FRW) universe.
In the present study, the effective energy and thermostatistical quantities on a cosmological horizon are examined to clarify the holographic-like connection between them, with a focus on a de Sitter universe.
To this end, the Helmholtz free energy on the horizon is derived from horizon thermodynamics.
The free energy is found to be equivalent to the effective energy calculated from the Friedmann equation. 
This consistency is interpreted as a kind of holographic-like connection.
To examine this connection, Padmanabhan's holographic equipartition law, which is related to the origin of spacetime dynamics, is applied to a de Sitter universe.
It is found that the law should lead to a holographic-like connection.
The holographic-like connection is considered to be a bridge between thermostatistical quantities on the horizon and in the bulk.
For example, cosmological equations for a flat FRW universe can be derived from horizon thermodynamics by accepting the connection as a viable scenario.
In addition, a thermal entropy equivalent to the Bekenstein--Hawking entropy is obtained from the Friedmann equation using the concept of a canonical ensemble in statistical physics.
The present study should provide new insight into the discussion of horizon thermodynamics and cosmological equations.

\end{abstract}


\maketitle

\section{Introduction} 
\label{Introduction}

Our Universe is expected to be well described as a Friedmann--Robertson--Walker (FRW) universe, whose cosmological equations provide a considerable amount of information \cite{Weinberg1,Roy1,Ryden1}.
The effective energy based on the energy density can be calculated from the Friedman equation. 
To explain the accelerated expansion of the late universe \cite{PERL1998_Riess1998_Planck2018}, astrophysicists have proposed various cosmological models \cite{Bamba1Nojiri1,HDE_review,Frusciante}, including lambda cold dark matter ($\Lambda$CDM), time-varying $\Lambda (t)$ cosmology \cite{FreeseOverduin,Nojiri2006etc,Sola_2009-2020}, bulk viscous cosmology \cite{Weinberg0,BarrowLima,BrevikNojiri}, and thermodynamic cosmology \cite{EassonCai,Basilakos1,Sheykhi1,Sheykhi2Karami,Koma4,Koma578,Koma6,Koma9}.
For these models, the effective energy can be calculated from the modified Friedman equation.
In addition, many of these models imply that our Universe should approach a de Sitter universe at least in the last stage; that is, they imply that our Universe behaves as an ordinary macroscopic system \cite{Pavon2013Mimoso2013}.
The de Sitter universe is considered to be in thermal equilibrium based on horizon thermodynamics \cite{GibbonsHawking1977}, which is related to black hole thermodynamics \cite{Bekenstein1,Hawking1}.
(The thermodynamics of the universe has been examined from various viewpoints \cite{Easther1,Gong00_01,Egan1,Bamba2018Pavon2019,deSitter_entropy,Saridakis20192021}.)

Thermodynamic cosmological models \cite{EassonCai,Basilakos1,Sheykhi1,Sheykhi2Karami,Koma4,Koma578,Koma6,Koma9} are based on not only black hole thermodynamics but also the holographic principle, which refers to bulk information stored on the horizon \cite{Hooft-Bousso}.
The concept of black hole thermodynamics has been applied to a cosmological horizon \cite{Jacob1995,Padma2010,Verlinde1,HDE,Padmanabhan2004,ShuGong2011,Padma2012AB,Cai2012Moradpour,HashemiWang,Sheykhia2018,Krishna2017,Krishna2019,Mathew2022,Koma10Koma11,Koma12,Koma14,Koma15,Koma16}.
The effective energy based on the energy density calculated from the Friedman equation is expected to be related to the total energy stored on the cosmological horizon.
However, the total energy is twice as large as the effective energy calculated from the Friedmann equation; that is, it is not equivalent to the effective energy, as previously reported \cite{Koma17}.
Cosmological equations have been extensively studied from a thermodynamics viewpoint \cite{EassonCai,Basilakos1,Sheykhi1,Sheykhi2Karami,Koma4,Koma578,Koma6,Koma9,Pavon2013Mimoso2013,Jacob1995,Padma2010,Verlinde1,HDE,Padmanabhan2004,ShuGong2011,Padma2012AB,Cai2012Moradpour,HashemiWang,Sheykhia2018,Krishna2017,Krishna2019,Mathew2022,Koma10Koma11,Koma12,Koma14,Koma15,Koma16}.
However, the relation between the effective energy and thermostatistical quantities on the horizon has not been examined in detail \cite{Koma17}.
It is worthwhile to clarify this relation because it is expected to be a bridge between cosmological equations and horizon thermodynamics.

Recently, we found that the effective energy calculated from the Friedman equation should be equivalent to the Helmholtz free energy on the horizon.
This holographic-like connection may allow us to develop a deeper understanding of thermodynamic scenarios based on the holographic principle.
For example, thermal entropy can be directly calculated from the free energy using the concept of a canonical ensemble in statistical physics \cite{Callen}.
We expect that Padmanabhan's holographic equipartition law \cite{Padma2012AB}, which is related to the origin of spacetime dynamics, is suitable for examining the holographic-like connection because
based on this law, cosmological equations are derived from the expansion of cosmic space due to the difference between the degrees of freedom on the surface and in the bulk \cite{Padma2012AB}.
(This law has been used to examine the emergence of cosmological equations \cite{Cai2012Moradpour,HashemiWang,Sheykhia2018,Krishna2017,Krishna2019,Mathew2022,Koma10Koma11,Koma12}.)
That is, Padmanabhan's holographic equipartition law is expected to be a bridge between thermostatistical quantities on the surface and in the bulk. 
An understanding of the relation between the effective energy calculated from the Friedman equation and thermostatistical quantities on the horizon should provide new insight into the discussion of cosmological equations and horizon thermodynamics.

In this context, we examine the effective energy calculated from the Friedman equation and the thermostatistical quantities on a cosmological horizon.
In the present study, we consider a flat FRW universe, with a focus on a de Sitter universe whose horizon is considered to be in thermal equilibrium.

The remainder of the present article is organized as follows.
In Sec.\ \ref{Cosmological equations}, cosmological equations are reviewed.
In addition, the effective energy based on the energy density is calculated from the Friedmann equation.
In Sec.\ \ref{Horizon thermodynamics}, the Helmholtz free energy on the horizon is derived from horizon thermodynamics.
In Sec.\ \ref{Padmanabhan's holographic equipartition law and an effective energy}, a holographic-like connection between the effective energy calculated from the Friedmann equation and the free energy on the horizon is examined through Padmanabhan's holographic equipartition law.
In Sec.\ \ref{Statistical physics}, thermostatistical quantities on the horizon are calculated from the Friedmann equation using the concept of a canonical ensemble in statistical physics.
Finally, in Sec.\ \ref{Conclusions}, the conclusions of the study are presented.

\section{Cosmological equations and effective energy} 
\label{Cosmological equations}

We consider a homogeneous, isotropic, and spatially flat universe, namely a flat FRW universe.
In Sec.\ \ref{A general formulation}, a general formulation for cosmological equations is introduced.
In Sec.\ \ref{Standard cosmological equations}, the standard cosmological equations used in this study are reviewed.
In Sec.\ \ref{Effective energy}, the effective energy based on the energy density is calculated from the Friedmann equation.

\subsection{General formulation} 
\label{A general formulation}
We introduce a general formulation for cosmological equations using the scale factor $a(t)$ at time $t$ according to previous works \cite{Koma6,Koma9,Koma14,Koma15,Koma16}.
The general Friedmann and acceleration equations are written as 
\begin{equation}
 H(t)^2      =  \frac{ 8\pi G }{ 3 } \rho (t)    + f_{\Lambda}(t)            ,                                                 
\label{eq:General_FRW01_f_0} 
\end{equation} 
\begin{align}
  \frac{ \ddot{a}(t) }{ a(t) }   &=  \dot{H}(t) + H(t)^{2}                                                                        \notag \\
                                          &=  -  \frac{ 4\pi G }{ 3 }  ( 1+  3w ) \rho (t)                                   +   f_{\Lambda}(t)    +  h_{\textrm{B}}(t)  ,  
\label{eq:General_FRW02_g_0}
\end{align}
where the Hubble parameter $H(t)$ is defined by
\begin{equation}
   H(t) \equiv   \frac{ da/dt }{a(t)} =   \frac{ \dot{a}(t) } {a(t)}  , 
\label{eq:Hubble}
\end{equation}
and $w$ represents the equation-of-state parameter for a generic component of matter, which is given as  
\begin{equation}
  w = \frac{ p(t) } { \rho(t)  c^2 }    .
\label{eq:w}
\end{equation}
Here, $G$, $c$, $\rho(t)$, and $p(t)$ are the gravitational constant, the speed of light, the mass density of cosmological fluids, and the pressure of cosmological fluids, respectively \cite{Koma6,Koma9,Koma14,Koma15,Koma16}.
For a $\Lambda$-dominated universe, a matter-dominated universe, and a radiation-dominated universe, $w$ is $-1$, $0$, and $1/3$, respectively.
The $\Lambda$-dominated universe corresponds to a de Sitter universe.
Two extra driving terms, $f_{\Lambda}(t)$ and $h_{\textrm{B}}(t)$, are phenomenologically assumed for a $\Lambda(t)$ cosmology and a bulk viscous cosmology, respectively \cite{Koma9,Koma16}.
The general formulation is discussed in Sec.\ \ref{Padmanabhan's holographic equipartition law}.

\subsection{Standard cosmological equations}
\label{Standard cosmological equations}

We now consider the standard cosmological equations.
To this end, we set $f_{\Lambda}(t) = h_{\textrm{B}}(t)=0$.
Consequently, Eqs.\ (\ref{eq:General_FRW01_f_0}) and (\ref{eq:General_FRW02_g_0}) are respectively rewritten as
\begin{equation}
 H^2      =  \frac{ 8\pi G }{ 3 } \rho              ,                                                 
\label{eq:FRW01} 
\end{equation} 
\begin{align}
  \frac{ \ddot{a} }{ a }  =  \dot{H} + H^{2}   = -  \frac{ 4\pi G }{ 3 }  ( 1+  3w ) \rho            .
\label{eq:FRW02}
\end{align}
In addition, the continuity equation is given by
\begin{equation}
       \dot{\rho} + 3  H (1+w)  \rho   =    0   .
\label{eq:drho}
\end{equation}
The continuity equation is derived from the first law of thermodynamics under the assumption of adiabatic processes \cite{Ryden1,Koma4}.
(Alternatively, the continuity equation can be derived from the Friedmann and acceleration equations because two of the three equations are independent \cite{Ryden1}.)

In the present paper, we focus on a de Sitter universe.
The evolution of the scale factor is given by 
\begin{equation}  
       a  =  a_{0}  \exp[ H ( t - t_{0} ) ]     ,
\label{eq:aa0_deSitter}
\end{equation}
where the subscript $0$ represents the present time.
The properties of a de Sitter universe are characterized by $H$, which is constant during the evolution of the de Sitter universe \cite{Koma17}.
In addition, $H$ is related to the temperature $T_{H}$ on the Hubble horizon, expressed as \cite{GibbonsHawking1977} 
\begin{equation}
 T_{H} = \frac{ \hbar H}{   2 \pi  k_{B}  }   ,
\label{eq:T_H}
\end{equation}
where $k_{B}$ and $\hbar$ are the Boltzmann constant and the reduced Planck constant, respectively. 
The reduced Planck constant is defined as $\hbar \equiv h/(2 \pi)$, where $h$ is the Planck constant \cite{Koma10Koma11,Koma12}.
Equation\ (\ref{eq:T_H}) indicates that $T_{H}$ depends only on $H$.
Accordingly, the horizon of a de Sitter universe is considered to be in thermal equilibrium because $T_{H}$ is constant.
That is, the concept of a canonical ensemble in statistical physics can be applied to the horizon of a de Sitter universe \cite{Koma17}.

\subsection{Effective energy $E_{\rm{eff}}$ based on energy density calculated from Friedmann equation} 
\label{Effective energy}

In this subsection, we determine the effective energy based on the energy density calculated from the Friedmann equation, which corresponds to the energy equation.
Using the Friedmann equation given by Eq.\ (\ref{eq:FRW01}), the energy density $\rho c^{2}$ is written as 
\begin{equation}
 \rho c^{2}  =    \frac{ 3 c^{2} }{ 8 \pi G }  H^{2}   .
\label{Energy-density_FRW}
\end{equation}
The energy density $\rho c^{2}$ can be generally defined by
\begin{equation}
 \rho c^{2}  =    \frac{ E_{\rm{eff}} }{ V }, 
\label{Energy-density_EeffV}
\end{equation}
where $E_{\rm{eff}}$ is the effective energy and $V$ is the volume of a sphere with the Hubble horizon (radius).
The Hubble horizon $r_{H}$ and the Hubble volume $V$ are respectively given by
\begin{equation}
     r_{H} = \frac{c}{H}   , 
\label{eq:rH}
\end{equation}
\begin{equation}
V = \frac{4 \pi}{3} r_{H}^{3} =  \frac{4 \pi}{3} \left ( \frac{c}{H} \right )^{3}   .
\label{eq:V}
\end{equation}
Note that the Hubble horizon is equivalent to an apparent horizon because of the flat universe.
Solving Eq.\ (\ref{Energy-density_EeffV}) with respect to $E_{\rm{eff}}$ and substituting Eqs.\ (\ref{Energy-density_FRW}) and (\ref{eq:V}) into the resultant equation yields 
\begin{align}
E_{\rm{eff}}  &= \rho c^{2} V =  \frac{ 3 c^{2} }{ 8 \pi G }  H^{2}    \frac{4}{3} \pi \left ( \frac{c}{H} \right )^3   = \frac{1}{2}   \frac{ c^{5} }{ G }  \left ( \frac{1}{H} \right )       .
\label{Eeff_FRW}
\end{align}

In this way, the effective energy $E_{\rm{eff}}$ based on the energy density can be calculated from the Friedmann equation.
The effective energy is expected to be related to the Helmholtz free energy $F_{H}$ on a cosmological horizon.
We examine this in the next section.
(The free energy corresponds to useful energy obtained from a closed thermodynamic system at a constant temperature.)

\section{Horizon thermodynamics} 
\label{Horizon thermodynamics}
Horizon thermodynamics is closely related to the holographic principle \cite{Hooft-Bousso}, which assumes that the horizon of the universe has an associated entropy and an approximate temperature \cite{EassonCai}.
In this section, based on horizon thermodynamics, the energy and the Helmholtz free energy on the horizon are derived. 
To this end, in Sec.\ \ref{Bekenstein-Hawking entropy}, the entropy and the temperature on the horizon are introduced according to previous works \cite{Jacob1995,Padma2010,Verlinde1,HDE,Padma2012AB,Cai2012Moradpour,HashemiWang,Sheykhia2018,Padmanabhan2004,ShuGong2011,Koma17}.
In Sec.\ \ref{Energy and free energy on the horizon}, the energy and the free energy on the horizon are derived using the entropy and the temperature.
In addition, the effective energy based on the energy density is discussed.
Note that we consider a de Sitter universe whose horizon is in thermal equilibrium.

\subsection{Definition of entropy $S_{H}$ and temperature $T_{H}$} 
\label{Bekenstein-Hawking entropy}

We select a form of the Bekenstein--Hawking entropy as an associated entropy on the cosmological horizon because it is the most standard, as examined in previous works \cite{Jacob1995,Padma2010,Verlinde1,HDE,Padma2012AB,Cai2012Moradpour,HashemiWang,Sheykhia2018,Padmanabhan2004,ShuGong2011,Koma17}.
In this subsection, we introduce the entropy on the horizon according to those works.

Based on the selected form of the Bekenstein--Hawking entropy, the entropy $S_{H}$ on the Hubble horizon is written as \cite{Koma17}
\begin{equation}
 S_{H}  = \frac{ k_{B} c^3 }{  \hbar G }  \frac{A_{H}}{4}   ,
\label{eq:SBH}
\end{equation}
where $A_{H}$ is the surface area of a sphere with the Hubble horizon.
Substituting $A_{H}=4 \pi r_{H}^2 $ into Eq.\ (\ref{eq:SBH}) and applying Eq.\ (\ref{eq:rH}) yields 
\begin{equation}
S_{H}  = \frac{ k_{B} c^3 }{  \hbar G }   \frac{A_{H}}{4}       
                  =  \left ( \frac{ \pi k_{B} c^5 }{ \hbar G } \right )  \frac{1}{H^2}  
                  =    \frac{K}{H^2}    , 
\label{eq:SBH2}      
\end{equation}
where $K$ is a positive constant given by
\begin{equation}
  K =  \frac{  \pi  k_{B}  c^5 }{ \hbar G } = \frac{  \pi  k_{B}  c^2 }{ L_{P}^{2} }   , 
\label{eq:K-def}
\end{equation}
and $L_{P}$ is the Planck length, written as
\begin{equation}
  L_{P} = \sqrt{ \frac{\hbar G} { c^{3} } }      .
\label{eq:Lp}
\end{equation}
The standard form given by Eq.\ (\ref{eq:SBH2}) is used for the entropy on the Hubble horizon.
(Forms of the non-Gaussian black-hole entropy have been described elsewhere \cite{Tsallis2012,Czinner1Czinner2,Barrow2020,Nojiri2021}.)

In addition, from Eq.\ (\ref{eq:T_H}), the temperature $T_{H}$ on the horizon is written as \cite{GibbonsHawking1977} 
\begin{equation}
 T_{H} = \frac{ \hbar H}{   2 \pi  k_{B}  }   .
\label{eq:T_H1}
\end{equation}
The horizon of a de Sitter universe is considered to be in thermal equilibrium because $T_{H}$ is constant.
Throughout this study, Eq.\ (\ref{eq:T_H1}) is used as an important thermostatistical quantity.

\subsection{Derivation of energy $E_{H}$ and free energy $F_{H}$} 
\label{Energy and free energy on the horizon}

In this subsection, we derive the energy $E_{H}$ stored on the Hubble horizon from the thermodynamic relation $dE_{H} =T_{H} dS_{H}$.
In addition, we derive the Helmholtz free energy $F_{H}$ on the horizon.
Hereafter, we refer to $F_{H}$ as the free energy (on the horizon).

Substituting Eq.\ (\ref{eq:T_H1})  and  $ d S_{H} =  \left (- \frac{2 K dH }{H^{3}} \right ) $ given by Eq.\ (\ref{eq:SBH2}) into $d E_{H} = T_{H} d S_{H}$ yields 
\begin{align}
 d E_{H} &= T_{H} d S_{H} =  \left ( \frac{ \hbar H}{   2 \pi  k_{B}  } \right )  \left (- \frac{2 K dH }{H^{3}} \right )  \notag \\
            &=  \frac{ \hbar }{   2 \pi  k_{B}  }   \left (- \frac{2 K  }{H^{2}} \right )   dH                                         .
\end{align}
Integrating this equation and using Eqs.\ (\ref{eq:SBH2}) and (\ref{eq:T_H1}) yields  
\begin{align}
 E_{H} &= \int  d E_{H} = \int \frac{ \hbar }{   2 \pi  k_{B}  }   \left (- \frac{2 K  }{H^{2}} \right ) dH  \notag \\
         &=  \frac{ \hbar }{   2 \pi  k_{B}  }    \frac{2K}{H}   + E_{H,0}    =  2 \left ( \frac{K}{H^{2}} \right ) \left ( \frac{ \hbar H}{   2 \pi  k_{B}  } \right )  + E_{H,0}    \notag \\
         &= 2 S_{H} T_{H} + E_{H,0} = 2 S_{H} T_{H}    ,
\label{E_ST2_thermo}
\end{align}
where the integral constant $E_{H,0}$ is set to zero so that $E_{H}=0$ is satisfied when $T_{H}=0$.
That is, the energy $E_{H}$ stored on the Hubble horizon is given by
\begin{align}
 E_{H} &=  2 S_{H} T_{H}    .
\label{E_ST2_thermo_2}
\end{align}
Substituting Eqs.\ (\ref{eq:SBH2}) and (\ref{eq:T_H1}) into Eq.\ (\ref{E_ST2_thermo_2}) yields
\begin{equation}
   E_{H}  =  2  \left [ \left ( \frac{ \pi k_{B} c^5 }{ \hbar G } \right )  \frac{1}{H^2}  \right ]  \left ( \frac{ \hbar H}{   2 \pi  k_{B}  }    \right )  =    \frac{ c^{5} }{ G }  \left ( \frac{1}{H} \right )  .
\label{E_ST2_thermo_H}
\end{equation}
In this way, $E_{H}= 2 S_{H} T_{H}$ and $E_{H} =  \frac{ c^{5} }{ G }  \left ( \frac{1}{H} \right )$ are obtained from $dE_{H} =T_{H} dS_{H}$.

In addition, based on thermodynamics, the free energy $F_{H}$ on the horizon is defined as
\begin{equation}
   F_{H} = E_{H} - T_{H} S_{H}  .
\label{F_def_thermo}
\end{equation}
Substituting $T_{H} S_{H} = E_{H}/2$ given by Eq.\ (\ref{E_ST2_thermo_2}) into Eq.\ (\ref{F_def_thermo}) yields
\begin{equation}
   F_{H} = E_{H} - T_{H} S_{H} = E_{H} - \frac{1}{2}E_{H} = \frac{1}{2}E_{H}  .
\label{F_2_thermo}
\end{equation}
The free energy $F_{H}$ is half of $E_{H}$.
Substituting Eq.\ (\ref{E_ST2_thermo_H}) into Eq.\ (\ref{F_2_thermo}) yields
\begin{equation}
   F_{H}  = \frac{1}{2}E_{H}  =   \frac{1}{2} \frac{ c^{5} }{ G }  \left ( \frac{1}{H} \right ) .
\label{F_3_thermo}
\end{equation}
As expected, $F_{H}$ is equivalent to $E_{\rm{eff}}$ given by Eq.\ (\ref{Eeff_FRW}).
That is, the free energy $F_{H}$ on the Hubble horizon is equivalent to the effective energy $E_{\rm{eff}}$ in the Hubble volume:
\begin{equation}
  F_{H}  = E_{\rm{eff}}     .
\label{F_Eeff_thermo}
\end{equation}
This consistency, namely Eq.\ (\ref{F_Eeff_thermo}), is considered to be a kind of holographic-like connection.
The interpretation of this connection is discussed in the next section using Padmanabhan's holographic equipartition law, which is related to the origin of spacetime dynamics \cite{Padma2012AB}.

When $F_{H}  = E_{\rm{eff}}$, the energy density $\rho c^{2}$ for the Hubble volume should be written as
\begin{align}
 \rho c^{2} &= \frac{ F_{H}   }{V}= \frac{E_{H}/2}{V}= \frac{ \frac{1}{2}   \frac{ c^{5} }{ G }  \left ( \frac{1}{H} \right )    }{\frac{4}{3} \pi (c/H)^3}    =   \frac{ 3 c^{2} }{ 8 \pi G }  H^{2}   .
\label{energy-density_F_thermo}
\end{align}
The above equation is equivalent to Eq.\ (\ref{Energy-density_FRW}), namely the Friedmann equation given by Eq.\ (\ref{eq:FRW01}).
That is, the Friedmann equation should be derived from the free energy on the horizon.
Also, it is well-known that the continuity equation given by Eq.\ (\ref{eq:drho}) is derived from the first law of thermodynamics \cite{Ryden1}.
Accordingly, the acceleration equations can be derived using the Friedmann and continuity equations because two of the three equations are independent.
In this way, the standard cosmological equations are derived from horizon thermodynamics when $F_{H}= E_{\rm{eff}}$.
(Note that another energy density obtained from $ E_{H}$ is given by $\frac{ E_{H} }{V} = \frac{ 3  c^{2} }{ 4 \pi G }  H^{2}$, as previously reported \cite{Koma17}. This is twice as large as Eq.\ (\ref{energy-density_F_thermo}), as described in Sec.\ \ref{Introduction}.) 

We expect that the holographic-like connection $F_{H}= E_{\rm{eff}}$ is a bridge between cosmological equations and horizon thermodynamics.
In the next section, the holographic-like connection is examined through Padmanabhan's holographic equipartition law \cite{Padma2012AB}.

In this section, we derived $E_{H}$ and $F_{H}$ from the thermodynamic relation $dE_{H} =T_{H} dS_{H}$ based on horizon thermodynamics.
The equipartition law of energy on the horizon was not used here.
(The law is described in Sec.\ \ref{Reformulation}.)
It has been reported that the equipartition law of energy on the horizon can lead to $E_{H}= 2 S_{H} T_{H}$ \cite{Padma2010,Padmanabhan2004,ShuGong2011} and $E_{H} =  \frac{ c^{5} }{ G }  \left ( \frac{1}{H} \right )$ \cite{Koma14,Koma17}.
The present result is consistent with previous works.
However, the free energy $F_{H}$ and the holographic-like connection $F_{H}  = E_{\rm{eff}}$ were not discussed in those works.

\section{Holographic equipartition law and effective energy} 
\label{Padmanabhan's holographic equipartition law and an effective energy}

In this section, we examine the effective energy that can be obtained from Padmanabhan's holographic equipartition law.
To this end, the law is introduced in Sec.\ \ref{Padmanabhan's holographic equipartition law} and its energy formula is discussed in Sec.\ \ref{Reformulation}.
In these two subsections, a flat FRW universe is considered.
In Sec.\ \ref{Effective energy in a de Sitter universe}, the energy formula is applied to a de Sitter universe to examine the effective energy.

\subsection{Padmanabhan's holographic equipartition law} 
\label{Padmanabhan's holographic equipartition law}

Based on Padmanabhan's holographic equipartition law, cosmological equations can be derived from the expansion of cosmic space due to the difference between the degrees of freedom on the surface and in the bulk in a region of space \cite{Padma2012AB}.
The law was reviewed in a previous study \cite{Koma10Koma11} based on Padmanabhan's work \cite{Padma2012AB} and other related studies \cite{Cai2012Moradpour,Sheykhia2018}.
In this subsection, we introduce Padmanabhan's holographic equipartition law based on those previous works.
Note that the law has not yet been established in a cosmological spacetime, but is considered to be a viable scenario.

In an infinitesimal interval $dt$ of cosmic time, the increase $dV$ in the Hubble volume can be expressed as 
\begin{equation}
     \frac{dV}{dt}  =  L_{p}^{2} (N_{\rm{sur}} - \epsilon N_{\rm{bulk}} ) \times c      , 
\label{dVdt_N-N}
\end{equation}
where $N_{\rm{sur}}$ is the number of degrees of freedom on a spherical surface of the Hubble radius $r_{H}$ and $N_{\rm{bulk}}$ is the number of degrees of freedom in the bulk \cite{Padma2012AB}. 
$L_{p}$ is the Planck length given by Eq.\ (\ref{eq:Lp}) and $\epsilon$ is a parameter defined as \cite{Padma2012AB}
 \begin{equation}
        \epsilon \equiv     
 \begin{cases}
              +1  & (\rho c^2 + 3p <0  \textrm{: an accelerating universe}),  \\ 
              -1  & (\rho c^2 + 3p >0   \textrm{: a decelerating universe}).    \\
\end{cases}
\label{epsilon}
\end{equation}
We refer to Eq.\ (\ref{dVdt_N-N}) as Padmanabhan's holographic equipartition law.
From this equation, the acceleration equation can be derived.
The derivation used in this subsection is based on previous reports \cite{Koma10Koma11}.
(Equation\ (\ref{dVdt_N-N}) includes $c$ because $c$ is not set to $1$.)

To derive the acceleration equation, we first calculate the left-hand side of Eq.\ (\ref{dVdt_N-N}), namely $dV/dt$. 
Differentiating Eq.\ (\ref{eq:V}) with respect to $t$ yields \cite{Koma10Koma11}
\begin{equation}
     \frac{dV}{dt}  =   \frac{d}{dt} \left ( \frac{4 \pi}{3} \left ( \frac{c}{H} \right )^{3}  \right )   =  -4 \pi c^{3}   \left (  \frac{ \dot{H} }{H^{4} } \right )  , 
\label{dVdt_right}
\end{equation}
where $r$ is set to $r_{H}=c/H$ before the time derivative is calculated \cite{Padma2012AB}.   
Next, to calculate the right-hand side of Eq.\ (\ref{dVdt_N-N}), the number of degrees of freedom in the bulk $N_{\rm{bulk}}$ is assumed to obey the equipartition law of energy \cite{Padma2012AB}: 
\begin{equation}
  N_{\rm{bulk}} = \frac{|E_{\rm{bulk}}|}{ \frac{1}{2} k_{B} T_{H}}     , 
\label{N_bulk}
\end{equation}
where the Komar energy $|E_{\rm{bulk}}|$ contained inside the Hubble volume $V$ is assumed to be given by 
\begin{equation}
|E_{\rm{bulk}}| =  |( \rho c^2 + 3p)| V  = - \epsilon ( \rho c^2 + 3p) V  .
\label{Komar}
\end{equation}
The Komar energy is different from the effective energy based on the energy density calculated from the Friedmann equation.
The number of degrees of freedom on the spherical surface $N_{\rm{sur}}$ is assumed to be given by  
\begin{equation}
  N_{\rm{sur}} = \frac{4 S_{\rm{sur}} }{k_{B}}       , 
\label{N_sur_P}
\end{equation}
where $S_{\rm{sur}}$ is the entropy on the horizon \cite{Koma10Koma11}.
For generality, $S_{\rm{sur}}$ is used here; $ S_{\rm{sur}} = S_{H}$ is used later.

We now derive an acceleration equation from Padmanabhan's holographic equipartition law.
According to previous works \cite{Padma2012AB,Koma10Koma11}, $\rho c^2 + 3p <0$ is selected and, therefore, $\epsilon = +1$ from Eq.\ (\ref{epsilon}).
This selection does not affect the following result \cite{Padma2012AB}.
We first calculate $N_{\rm{bulk}}$ on the right-hand side of Eq.\ (\ref{dVdt_N-N}).
Substituting Eqs.\ (\ref{eq:T_H1}) and (\ref{Komar}) into Eq.\ (\ref{N_bulk}) and using Eqs.\ (\ref{eq:w}) and (\ref{eq:V}) and $\epsilon = +1$ yields \cite{Koma10Koma11}
\begin{align}
  N_{\rm{bulk}}  &= \frac{|E_{\rm{bulk}}|}{ \frac{1}{2} k_{B} T_{H}}  
                       =   -  \frac{ (4 \pi)^{2} c^{5}  }{ 3 \hbar }  (1+3w)\rho  \frac{1}{  H^{4}   }      .
\label{N_bulk_cal}
\end{align}
In addition, substituting $\epsilon = +1$ and Eqs.\ (\ref{eq:Lp}), (\ref{dVdt_right}), (\ref{N_sur_P}), and (\ref{N_bulk_cal}) into Eq.\ (\ref{dVdt_N-N}) and solving the resultant equation with respect to $\dot{H}$ yields \cite{Koma10Koma11}  
\begin{align}
  \dot{H}                                                     
                  &=   -  \frac{ 4 \pi G }{ 3} (1+3w)  \rho   - \frac{ S_{\rm{sur}} H^{4} }{K}                        , 
\label{dVdt_N-N_cal2}
\end{align}
where $K$ is given by Eq.\ (\ref{eq:K-def}).
Substituting Eq.\ (\ref{dVdt_N-N_cal2}) into $ \ddot{a}/ a   =  \dot{H} + H^{2}$ and using $S_{H} = K/H^{2}$ given by Eq.\ (\ref{eq:SBH2}), we obtain the acceleration equation, written as \cite{Koma10Koma11}
\begin{align}
  \frac{ \ddot{a} }{ a }       &=  \dot{H} + H^{2}             
                                      =   -  \frac{ 4 \pi G }{ 3}  (1+3w)  \rho - \frac{ S_{\rm{sur}} H^{4} }{K}      + H^{2}  \notag \\
                                    &=   -  \frac{ 4 \pi G }{ 3}  (1+3w)  \rho + H^{2}  \left (  1 - \frac{ S_{\rm{sur}} }{ S_{H} }  \right )     .
\label{N-N_FRW02_SH}
\end{align}
When $S_{\rm{sur}} \neq S_{H}$, the second term $H^{2}(1- S_{\rm{sur}}/S_{H})$ on the right-hand side is non-zero \cite{Koma10Koma11}.
In this case, the obtained acceleration equation corresponds to the general form given by Eq.\ (\ref{eq:General_FRW02_g_0}).

When $ S_{\rm{sur}} = S_{H}$, the second term on the right-hand side of Eq.\ (\ref{N-N_FRW02_SH}) is zero and, therefore, the acceleration equation is written as 
\begin{align}
  \frac{ \ddot{a} }{ a }       &=  \dot{H} + H^{2}     =   -  \frac{ 4 \pi G }{ 3}  (1+3w)  \rho     .
\label{N-N_FRW02_SH_00}
\end{align}
This equation is equivalent to Eq.\ (\ref{eq:FRW02}) in standard cosmology.
For a specific case, we consider $dV/dt = 0$, corresponding to a de Sitter universe \cite{Padma2012AB}.
That is, $\dot{H}=0$ and $w=-1$ are considered.
Substituting $\dot{H}=0$ and $w=-1$ into Eq.\ (\ref{N-N_FRW02_SH_00}), we have the Friedmann equation, written as \cite{Padma2012AB}
\begin{equation}
 H^2      =  \frac{ 8\pi G }{ 3 } \rho              .                                                
\label{eq:FRW01_2} 
\end{equation} 
The specific case is discussed again in Sec.\ \ref{Effective energy in a de Sitter universe} to examine the effective energy of a de Sitter universe.
To this end, in the next subsection, we derive an energy formula of Padmanabhan's holographic equipartition law.

\subsection{Energy formula of Padmanabhan's holographic equipartition law}
\label{Reformulation}

In this subsection, we derive an energy formula of Padmanabhan's holographic equipartition law.
We have considered $\rho c^2 + 3p <0$ and $\epsilon = +1$.
Accordingly, Eq.\ (\ref{dVdt_N-N}) is rewritten as
\begin{equation}
     \frac{dV}{dt}  =  L_{p}^{2} (N_{\rm{sur}} -  N_{\rm{bulk}} ) \times c      . 
\label{dVdt_N-N_2}
\end{equation}
We first calculate the left-hand side of Eq.\  (\ref{dVdt_N-N_2}), namely $dV/dt$. 
To this end, Eq.\ (\ref{dVdt_right}) and $\dot{S}_{H} =\frac{-2K \dot{H} }{H^{3}}$ are used.
From $\dot{S}_{H} =\frac{-2K \dot{H} }{H^{3}}$, Eq.\ (\ref{dVdt_right}) can be rewritten as
\begin{equation}
     \frac{dV}{dt}   =  -4 \pi c^{3}   \left (  \frac{ \dot{H} }{H^{4} } \right )  =  \frac{ \frac{ -2 K \dot{H} }{ H^{3} }   }{  H  }  \frac{ 2 \pi c^{3} }{K}  =  \frac{ \dot{S}_{H}  }{ H }  \frac{ 2 \pi c^{3} }{K}   .
\label{dVdt_right_dotS_0}
\end{equation}
Substituting Eq.\ (\ref{dVdt_right_dotS_0}) into Eq.\ (\ref{dVdt_N-N_2}) yields
\begin{equation}
     \frac{ \dot{S}_{H}  }{ H }  \frac{ 2 \pi c^{3} }{K}  =  L_{p}^{2}  (N_{\rm{sur}} -  N_{\rm{bulk}} ) \times c    .
\label{dVdt_N-N_3}
\end{equation}
Solving the above equation with respect to $\dot{S}_{H}$ and using  $K =  \frac{  \pi  k_{B}  c^2 }{ L_{P}^{2} }$ given by Eq.\ (\ref{eq:K-def}) yields
\begin{align}
     \dot{S}_{H}  &= \frac{H K}{2 \pi c^{3} } L_{p}^{2}  (N_{\rm{sur}} -  N_{\rm{bulk}} ) \times  c                                                  \notag \\
                      &=\frac{H  \frac{  \pi  k_{B}  c^2 }{ L_{P}^{2} } }{2 \pi c^{3} } L_{p}^{2}  (N_{\rm{sur}} -  N_{\rm{bulk}} ) \times  c   \notag \\
                      &=\frac{k_{B}}{2} H   (N_{\rm{sur}} -  N_{\rm{bulk}} )   .
\label{dSH_N-N}
\end{align}
This equation was derived and examined by Krishna and Mathew \cite{Krishna2019}.

In addition, we set $ S_{\rm{sur}} = S_{H}$ for standard cosmology.
From Eq.\ (\ref{N_sur_P}), $N_{\rm{sur}}$ is given by
\begin{equation}
  N_{\rm{sur}} = \frac{4 S_{\rm{sur}} }{k_{B}}  =   \frac{4 S_{H} }{k_{B}}   .
\label{N_sur_N_H}
\end{equation}
Substituting $S_{H} =\frac{E_{H}}{2 T_{H}}$ given by Eq.\ (\ref{E_ST2_thermo_2}) into Eq.\ (\ref{N_sur_N_H}) yields
\begin{align}
  N_{\rm{sur}}  &=  \frac{4 S_{H} }{k_{B}} =   \frac{4  \left ( \frac{E_{H}}{2 T_{H}} \right ) }{k_{B}} =  \frac{ E_{H} }{ \frac{1}{2} k_{B} T_{H}}  .
\label{N_sur_E_H}
\end{align}
Equation\ (\ref{N_sur_E_H}) represents the equipartition law of energy on the horizon \cite{Padmanabhan2004,Padma2010}.
From Eq.\ (\ref{N_bulk}), the number of degrees of freedom in the bulk $N_{\rm{bulk}}$ is written as
\begin{equation}
  N_{\rm{bulk}} = \frac{|E_{\rm{bulk}}|}{ \frac{1}{2} k_{B} T_{H}}    .
\label{N_bulk_2}
\end{equation}
We note that the temperature $T_{H}$ on the horizon is included in Eqs.\ (\ref{N_sur_E_H}) and (\ref{N_bulk_2}).
Substituting Eqs.\ (\ref{N_sur_E_H}) and (\ref{N_bulk_2}) into Eq.\ (\ref{dSH_N-N}) yields
\begin{align}
     \dot{S}_{H}  &= \frac{k_{B}}{2} H   (N_{\rm{sur}} -  N_{\rm{bulk}} )                                                  
                      =  \frac{k_{B}}{2} H  \left (  \frac{ E_{H} }{ \frac{1}{2} k_{B} T_{H}}    -   \frac{|E_{\rm{bulk}}|}{ \frac{1}{2} k_{B} T_{H}}   \right )  \notag \\
                      &=   H   \left (   \frac{ E_{H} -  |E_{\rm{bulk}}| }{ T_{H}}       \right )   .
\label{dSH_E-E}
\end{align}
This is the energy formula of Padmanabhan's holographic equipartition law.
In the next subsection, the energy formula is applied to a de Sitter universe.
(The Komar energy $|E_{\rm{bulk}}|$ is different from the effective energy $E_{\rm{eff}}$ based on the energy density calculated from the Friedmann equation.)

\subsection{Effective energy $\tilde{E}_{\rm{eff}}$ in de Sitter universe calculated from Padmanabhan's holographic equipartition law}
\label{Effective energy in a de Sitter universe}

We consider a de Sitter universe and examine the effective energy calculated from Padmanabhan's holographic equipartition law.
A de Sitter universe corresponds to $\dot{S}_{H}=0$ because $H$ is constant.
Accordingly, $E_{H} = |E_{\rm{bulk}}|$ is obtained from Eq.\ (\ref{dSH_E-E}).
That is, $E_{H} = |E_{\rm{bulk}}|$ is considered to be the so-called holographic duality based on the law.

We now discuss the effective energy.
Substituting Eq.\ (\ref{Komar}) into $E_{H} = |E_{\rm{bulk}}|$ and using $w = p/(\rho c^2)$  given by Eq.\ (\ref{eq:w}) yields
\begin{align}
      E_{H} &=  |E_{\rm{bulk}}| =   |( \rho c^2 + 3p)| V =  \rho c^2 | 1 + 3w | V            ,
\label{E_E_deSitter}
\end{align}
where $\rho$ is assumed to be non-negative.
The equation-of-state parameter $w$ is retained here and set to $-1$ later.
($\epsilon$ given by Eq.\ (\ref{epsilon}) is not used here.)
Using Eq.\ (\ref{E_E_deSitter}), the energy density $\rho c^2$ is written as
\begin{align}
     \rho c^2   &=  \frac{E_{H}}{ | 1 + 3w | V }   =    \frac{\tilde{E}_{\rm{eff}}}{ V }                     ,
\label{rho_deSitter}
\end{align}
where $\tilde{E}_{\rm{eff}}$ is the effective energy in the Hubble volume and is defined as a new parameter, which is given by 
\begin{align}
    \tilde{E}_{\rm{eff}}   &=  \frac{E_{H}}{ | 1 + 3w | }     =    \frac{E_{H}}{ 2 }                 .
\label{Eeff_deSitter}
\end{align}
Here, $w=-1$ is used.
From Eq.\ (\ref{Eeff_deSitter}), we can confirm that $\tilde{E}_{\rm{eff}}$ is equivalent to both $F_{H}$ given by Eq.\ (\ref{F_3_thermo}) and $E_{\rm{eff}}$ given by Eq.\ (\ref{Eeff_FRW}):
\begin{align}
    \tilde{E}_{\rm{eff}}  &=  \frac{E_{H}}{2}  = F_{H}  = E_{\rm{eff}}   .                                
\label{Eeff_deSitter_Eeff_FRW_FH}
\end{align}
This result implies that Padmanabhan's holographic equipartition law should lead to a holographic-like connection $F_{H}= E_{\rm{eff}}$.
We note that $E_{H}$ and $|E_{\rm{bulk}}|$ are equivalent to $2 \tilde{E}_{\rm{eff}}$ because $E_{H} = |E_{\rm{bulk}}|$.

As mentioned above, $E_{H} = |E_{\rm{bulk}}|$ is considered to be the holographic duality.
In addition, $|E_{\rm{bulk}}|$ is equivalent to $2 \tilde{E}_{\rm{eff}}$.
Consequently, a holographic-like connection $F_{H}= E_{\rm{eff}} (=\tilde{E}_{\rm{eff}})$ given by Eq.\ (\ref{Eeff_deSitter_Eeff_FRW_FH}) is obtained.
Accordingly, the connection should be interpreted as a kind of reformulated holographic duality based on Padmanabhan's holographic equipartition law.

The holographic-like connection is expected to be a bridge between thermostatistical quantities on the horizon and in the bulk.
This connection can lead to thermostatistical quantities such as a thermal entropy using a canonical ensemble in statistical physics.
We examine this in the next section.

Note that an energy balance relation $\rho c^2 V =T_{H} S_{H}$, similar to the holographic-like connection, has been derived by Padmanabhan \cite{Pad2017}.
The energy balance relation was also described in, e.g., Ref.\ \cite{Tu20182019}.
However, the free energy $F_{H}$ and the holographic-like connection $F_{H}  = E_{\rm{eff}}$ were not discussed in those previous works.

\section{Statistical physics and Friedmann equation}
\label{Statistical physics}

So far, we have examined the free energy $F_{H}$ on the horizon and the effective energy $E_{\rm{eff}}$ based on the energy density calculated from the Friedmann equation.
A holographic-like connection $F_{H}= E_{\rm{eff}}$ was obtained.
This connection is related to Padmanabhan's holographic equipartition law and is expected to be a bridge between thermostatistical quantities on the horizon and in the bulk.
To confirm this expectation, we calculate thermostatistical quantities on the horizon from the Friedmann equation by accepting the holographic-like connection as a viable scenario and using statistical physics.
In Sec.\ \ref{Canonical ensemble in statistical physics}, a standard formulation for a canonical ensemble in statistical physics is introduced.
In Sec.\ \ref{Energy and Entropy on the horizon}, the formulation is applied to the horizon of a de Sitter universe to calculate the thermostatistical quantities on the horizon.

\subsection{Canonical ensemble in statistical physics}
\label{Canonical ensemble in statistical physics}

A standard formulation for a canonical ensemble is introduced \cite{Callen}.
We consider a canonical ensemble with a partition function $Z(\beta)$ given by 
\begin{equation}
Z (\beta) = \rm{Tr} \left [ e^{- \beta \mathcal{H} }  \right ]    ,
\label{Partition function}
\end{equation}
where $\mathcal{H}$ is the Hamiltonian for the system and $\beta$ is the inverse temperature in the equilibrium state, written as
\begin{equation}
\beta  = \frac{1}{k_{B} T} , 
\label{eq_beta}
\end{equation}
where $T$ represents the equilibrium temperature of the system.
Based on statistical physics, the free energy $F(\beta) $ is given by \cite{Callen}
\begin{equation}
F (\beta) = - \frac{1}{\beta} \ln Z (\beta) . 
\label{Free energy}
\end{equation}
Using $F(\beta) $, the total energy $E(\beta)$ is written as \cite{Callen}
\begin{align}
E (\beta) &=     -  \frac{\partial}{ \partial \beta }   \ln Z (\beta) = \frac{\partial}{ \partial \beta }  \beta F (\beta)   ,
\label{Total energy}
\end{align}
and the thermal entropy $S(\beta)$ is given by
\begin{equation}
\frac{ S(\beta)}{k_{B}} =   \ln Z(\beta) +   \beta E(\beta)   = \beta^{2} \frac{\partial}{\partial \beta}  F (\beta)    .
\label{Thermal entropy}
\end{equation}
Equations\ (\ref{Total energy}) and (\ref{Thermal entropy}) indicate that the total energy $E$ and the thermal entropy $S$ can be calculated from the free energy $F$.
In the next subsection, the formulations are applied to the horizon of a de Sitter universe.

\subsection{Thermal entropy derived from effective energy calculated from Friedmann equation}
\label{Energy and Entropy on the horizon}

In this subsection, based on statistical physics, we calculate the thermal entropy on the horizon from the Friedmann equation.
To this end, the formulations examined in Sec.\ \ref{Canonical ensemble in statistical physics} are applied to a de Sitter universe whose horizon is considered to be in thermal equilibrium.
The inverse temperature in the equilibrium state is written as
\begin{equation}
 \beta  = \frac{1}{k_{B} T}  = \frac{1}{k_{B} T_{H}  } , 
\label{eq_beta_H}
\end{equation}
where the temperature $T_{H}$ on the horizon is used.

In addition, we accept the holographic-like connection, namely $F_{H}=E_{\rm{eff}}$, as a viable scenario.
($E_{\rm{eff}}$ is the effective energy calculated from the Friedmann equation.)
Substituting Eq.\ (\ref{Eeff_FRW}) into $F_{H}=E_{\rm{eff}}$ yields 
\begin{align}
 F_{H} &= E_{\rm{eff}} = \frac{1}{2}   \frac{ c^{5} }{ G }  \left ( \frac{1}{H} \right )   .
\label{F_Eeff_FRW}
\end{align}
To discuss a canonical ensemble, the free energy $F_{H}$ is formulated as a function of the inverse temperature $\beta$.
Substituting $\frac{1}{H} = \frac{ \hbar }{ 2 \pi  k_{B} T_{H} }$ given by Eq.\ (\ref{eq:T_H1}) into Eq.\ (\ref{F_Eeff_FRW}) and using Eq.\ (\ref{eq_beta_H}) yields 
\begin{align}
 F_{H} &= \frac{1}{2}   \frac{ c^{5} }{ G } \left ( \frac{ \hbar }{ 2 \pi  k_{B} T_{H} }  \right )  =  \left ( \frac{ \hbar  c^{5}  }{ 4 \pi  G}  \right ) \beta . 
\label{F_Eeff_FRW_beta}
\end{align}

We now calculate the total energy $E$ and the thermal entropy $S$.
The symbols $E_{H}$ and $S_{H}$ are not used here because from now on, the total energy $E$ and the thermal entropy $S$ are calculated.
Substituting Eq.\ (\ref{F_Eeff_FRW_beta}) into Eq.\ (\ref{Total energy}) yields 
\begin{align}
E &=  \frac{\partial}{ \partial \beta }  \beta F_{H}  = \frac{\partial}{ \partial \beta }  \beta  \left [  \left ( \frac{ \hbar  c^{5}  }{ 4 \pi  G}  \right )  \beta    \right ]     
     =  \left ( \frac{ \hbar  c^{5}  }{ 2 \pi  G}  \right ) \beta =  2 F_{H}  . 
\label{Total energy_EH1}
\end{align}
From Eqs.\ (\ref{F_Eeff_FRW}) and (\ref{Total energy_EH1}), the total energy $E$ is written as
\begin{align}
E &=  \frac{ c^{5} }{ G }  \left ( \frac{1}{H} \right )   . 
\label{Total energy_EH2}
\end{align}
Substituting Eq.\ (\ref{F_Eeff_FRW_beta}) into Eq.\ (\ref{Thermal entropy}) yields 
\begin{align}
\frac{ S}{k_{B}} =  \beta^{2} \frac{\partial}{\partial \beta}  F_{H} = \beta^{2} \frac{\partial}{\partial \beta} \left [ \left ( \frac{ \hbar  c^{5}  }{ 4 \pi  G}  \right ) \beta  \right ]  =  \left ( \frac{ \hbar  c^{5}  }{ 4 \pi  G}  \right ) \beta^{2} .
\label{Thermal entropy_SH1}
\end{align}
From this equation, the thermal entropy $S$ is written as
\begin{align}
S      &=  k_{B}  \left ( \frac{ \hbar  c^{5}  }{ 4 \pi  G}  \right ) \beta^{2} =   \frac{ \hbar k_{B} c^5 }{ 4 \pi G }  \left ( \frac{1}{ k_{B}T_{H} } \right )^{2}                               \notag \\
       &=\left ( \frac{ \pi k_{B} c^5 }{ \hbar G } \right ) \left ( \frac{\hbar}{ 2 \pi  k_{B}T_{H} } \right )^{2}  =\left ( \frac{ \pi k_{B} c^5 }{ \hbar G } \right )  \frac{1}{H^2}     \notag \\
       &=\frac{ k_{B} c^3 }{  \hbar G }   \frac{4 \pi (c/H)^2 }{4}  =\frac{ k_{B} c^3 }{  \hbar G }   \frac{4 \pi r_H^2 }{4}    =\frac{ k_{B} c^3 }{  \hbar G }   \frac{A_{H}}{4}                        ,
\label{Thermal entropy_SH2}
\end{align}
where Eq.\ (\ref{eq_beta_H}) and $\frac{\hbar}{2 \pi  k_{B} T_{H} } =  \frac{1}{H}$ given by Eq.\ (\ref{eq:T_H1}) are used.
Substituting Eq.\ (\ref{eq:T_H1}) and the second line of Eq.\ (\ref{Thermal entropy_SH2}) into $2 S T_{H}$ and using Eq.\ (\ref{Total energy_EH2}) yields 
\begin{equation}
 2 S T_{H} = 2 \left ( \frac{ \pi k_{B} c^5 }{ \hbar G } \right )  \frac{1}{H^2}   \left ( \frac{ \hbar H}{   2 \pi  k_{B}  }  \right ) = \frac{ c^{5} }{ G }  \left ( \frac{1}{H} \right ) = E      .
\label{eq:2SH_TH_thermal}
\end{equation}

In this way, based on statistical physics, the total energy $E$ and the thermal entropy $S$ on the horizon are obtained from the effective energy $E_{\rm{eff}}$ calculated from the Friedmann equation.
We can confirm that the obtained equations, namely $E =2 F_{H}$, $E =  \frac{ c^{5} }{ G }  \left ( \frac{1}{H} \right )$, $S =\frac{ k_{B} c^3 }{  \hbar G }   \frac{A_{H}}{4}$, and $E =2 S T_{H} $, are equivalent to those examined in Sec.\ \ref{Horizon thermodynamics}.
The interpretation of the holographic-like connection $F_{H}= E_{\rm{eff}}$ has not yet been established.
However, when this connection is acceptable, the total energy and the thermal entropy on the horizon can be obtained from the Friedmann equation.
Using an inverse operation, the Friedmann equation can be calculated from the thermostatistical quantities on the horizon.
(A similar inverse operation was discussed in Sec.\ \ref{Horizon thermodynamics}.)
In this sense, the holographic-like connection is considered to be a bridge between horizon thermodynamics and cosmological equations (i.e., a dictionary between thermostatistical quantities on the horizon and in the bulk).
These results imply that thermostatistical quantities on the horizon and in the bulk can be examined through the holographic-like connection and Padmanabhan's holographic equipartition law.

In the present study, we assume a standard FRW cosmology, with a focus on a de Sitter universe in which $H$, $r_{H}$, and $T_{H}$ are constant.
The temperature $T_{H}$ given by Eq.\ (\ref{eq:T_H1}) is obtained from field theory in the de Sitter space \cite{GibbonsHawking1977}.
In addition, we select a form of the Bekenstein--Hawking entropy as an associated entropy on the cosmological horizon.
However, the selected from of the entropy on the horizon may modify the standard FRW cosmology.
In fact, modified FRW cosmologies have been examined based on various forms of the entropy, see, e.g., the recent work of Nojiri \textit{et al}. \cite{ApparentHorizon2022}.
Therefore, the question arises whether the holographic-like connection can be applied to the modified FRW cosmology.
Probably, it should be difficult to apply this connection directly to modified FRW universes because horizons of these universes are generally dynamic, unlike for the de Sitter universe whose horizon is considered to be static.
Instead of $T_{H}$, we may need to use a dynamical temperature, to extend the holographic-like connection.

In fact, a dynamical temperature has been proposed to describe the temperature on a dynamic apparent horizon \cite{Dynamical-T-20072014}.  
The dynamical temperature should be suitable for discussing thermodynamics and statistical physics on dynamic horizons.
Recently, we examined the dynamical temperature of a flat FRW universe, using a $\Lambda(t)$ model similar to time-varying $\Lambda(t)$ cosmologies \cite{Koma19}.
Consequently, we found a universe whose dynamical temperature is constant although $H$ and $r_{H}$ vary with time \cite{Koma19}.
Such a universe at constant dynamical-temperature is expected to contribute to the study of the holographic-like connection in modified FRW cosmologies because the temperature on the horizon is considered to be constant. 
We expect that an extended holographic-like connection obtained from the modified Friedmann equation can be discussed, by focusing on a specific universe at constant dynamical-temperature in the modified FRW cosmology.
Detailed studies are needed and are left for future research.

We note that the holographic entanglement entropy \cite{RyuTakayanagi2006}, which is related to the emergence of spacetime, has been extensively examined \cite{Takayanagi2007,Takayanagi2022_Page,Jacobson,Hartman2020,Takayanagi2022,deSitter,deSitter2,Nakaguchi,Relative_entropy,deSitter_Fluc,Verlinde2020,Zurek2021_2022} based on anti-de Sitter/conformal field theory \cite{Maldacena1997}.
The entanglement entropy can be treated as a thermal entropy; this has been examined in a de Sitter space \cite{Takayanagi2022,deSitter,deSitter2}. 
In particular, it has been reported that the formula for the entropy on a cosmological horizon in a de Sitter space should be equivalent to a holographic entanglement entropy \cite{deSitter2}.
In addition, the relative entropy, corresponding to the free-energy difference, is expected to be related to gravity \cite{Relative_entropy}. 
We may be able to discuss a universe from a new viewpoint using these entropies and the holographic-like connection examined in the present paper.
This task is left for future research.

\section{Conclusions}
\label{Conclusions}

We examined the effective energy $E_{\rm{eff}}$ based on the energy density calculated from the Friedmann equation and thermostatistical quantities on a cosmological horizon of a de Sitter universe to clarify the holographic-like connection between the two.

First, we derived the energy $E_{H}$ and the Helmholtz free energy $F_{H}$ on the horizon from horizon thermodynamics.
The free energy $F_{H}$ was found to be half of the energy $E_{H}$ on the horizon and equivalent to the effective energy $E_{\rm{eff}}$ obtained using the Friedmann equation, that is, $E_{\rm{eff}} = F_{H} = E_{H}/2$.
The consistency is considered to be a kind of holographic-like connection.
When this connection is acceptable, the Friedmann equation can be derived from the free energy $F_{H}$ on the horizon, that is, cosmological equations for a flat FRW universe can be derived from horizon thermodynamics.

Second, we examined the holographic-like connection through Padmanabhan's holographic equipartition law.
Applying the law to a de Sitter universe, an effective energy $\tilde{E}_{\rm{eff}}$ in the Hubble volume was found to be given by $E_{H}/2$, that is, $\tilde{E}_{\rm{eff}} = \frac{E_{H}}{2} =  F_{H} =E_{\rm{eff}}$.
This result implies that Padmanabhan's holographic equipartition law is related to the holographic-like connection $F_{H}= E_{\rm{eff}}$ and should lead to the connection.
The holographic-like connection can be interpreted as a kind of reformulated holographic duality based on the law.

Finally, we examined thermostatistical quantities on the horizon by accepting the connection $F_{H}= E_{\rm{eff}}$ as a viable scenario and using the concept of a canonical ensemble in statistical physics.
We confirmed that a thermal entropy equivalent to the form of the Bekenstein--Hawking entropy is obtained from the Friedmann equation.
Using an inverse operation, the Friedmann equation can be calculated from thermostatistical quantities on the horizon.

The holographic-like connection is considered to be a bridge between horizon thermodynamics and cosmological equations (i.e., a dictionary between thermostatistical quantities on the horizon and in the bulk).
The present study should provide new insight into the discussion of gravity and horizon thermodynamics.

\appendix

\end{document}